\documentclass{PoS}

\title{The equation of state at high temperatures from lattice QCD}

\ShortTitle{The equation of state at high temperatures from lattice QCD}

\author{G.~Endr\H{o}di$^1$, Z. Fodor$^{1,2}$, S.D.~Katz$^{1,2}$, \speaker{K.K. Szab\'o}$^2$\\
$^1$Institute for Theoretical Physics, E\"otv\"os University, H-1117 Budapest, Hungary. \\
$^2$Department of Physics, University of Wuppertal, D-42097 Wuppertal, Germany.\\
       E-mail: \email{szaboka@general.elte.hu}
}

\abstract{We present results for the equation of state upto previously unreachable, high temperatures. Since the temperature range is quite large, a comparison with perturbation theory can be done directly. 
}

\FullConference{The XXV International Symposium on Lattice Field Theory\\
                July 30 - August 4 2007\\
                Regensburg, Germany}

\begin{document}

\section{Introduction}

Quantum chromodynamics (QCD) is the theory describing the strong interactions. According to QCD, as the temperature increases, hadronic matter undergoes a transition to quark gluon plasma. The function, which describes the equilibrium in this system for different temperatures, is the equation of state. For the equation of state, practically one has to measure only the pressure, which then will unambiguously determine the energy density as well. A difficulty emerging about the equation of state is, that perturbation theory does not seem to fit lattice results even at higher temperatures (smaller couplings). While available lattice results both for pure gauge theory~\cite{Boyd:1996bx,Okamoto:1999hi} and for full QCD~\cite{Karsch:2000ps, Bernard:2006nj, Aoki:2005vt} (and see also~\cite{Karsch:lat07}) end at around $5 \cdot T_C$, standard perturbation theory converges only at extremely high temperatures. To create a link between these two methods, we present lattice results on the pressure at temperatures, which were previously unreachable. It becomes possible to compare our data with perturbation theory formulae. Our results are obtained using two different approaches: we present a new way to renormalize the pressure (for this we have results for $N_t=4$), and a direct method to measure the pressure (results for $N_t=4,6$ and $8$).

\section{Renormalization of the pressure}

For illustration, we present here the technique and the results for pure SU(3) gauge theory. The extension to full QCD is straightforward.

In order to obtain results which can be extrapolated to the continuum, one has to renormalize the pressure. This is usually done using the standard integral method~\cite{Engels:1990vr}:

\begin{equation}
p_{\rm ren}(T)=p(T) - p(0)=\int d\beta \left( \langle {\rm Pl} \rangle_{T}-\langle {\rm Pl} \rangle_{0} \right)
\label{eq:p_int}
\end{equation}
This way one has to carry out simulations on finite and zero temperature lattices. However, exactly zero temperature cannot be realized. Together with the fact, that divergences that are removed by renormalization are independent of the temperature, rises the question: why not use finite $T$ lattices for renormalization? In order to do so, let us introduce the following quantity:

\begin{equation}
\bar{p}(T)= p(T)-p(T/2)=\int d\beta \left( \langle {\rm Pl} \rangle_{N_t}-\langle {\rm Pl} \rangle_{2\cdot N_t} \right)
\label{eq:p_bar}
\end{equation}
Here, instead of using $1/2$ for the subtraction temperature, one can use any factor, which is smaller than $1$. Now, we may build up $p_{\rm ren}$ as a sum of differences, so it can be expressed with $\bar p$ as:

\begin{equation}
p_{\rm ren}(T)=p(T)-p(T/2)+p(T/2)-p(T/4)+\dots=\bar{p}(T)+\bar{p}(T/2)+\dots
\label{eq:p_sum}
\end{equation}
In fact, we usually measure the dimensionless pressure, which can be obtained by including the $T^4$ factors:

\begin{equation}
\frac{p_{\rm ren}}{T^4}=\left.\frac{\bar{p}}{T^4}\right|_T+\left.\frac{1}{16}\cdot \frac{\bar{p}}{T^4}\right|_{T/2}+\frac{1}{256}\cdot\dots
\end{equation}
Due to the increasing powers of $1/16$ in the forthcoming terms, one practically needs only a few of them. Applying this scheme, we may reach arbitrary high temperatures using lattices with only $N_t$ and $2 \cdot N_t$ temporal extent. It is worth mentioning, that a similar formula can be constructed for the case of the normalized interaction measure $I \equiv (\epsilon - 3 \cdot p) / T^4$:

\begin{equation}
I_{\rm ren}(T)=\bar{I}(T)+\frac{1}{16}\bar{I}(T/2)+\dots
\end{equation}
where $$\bar{I}(T)=N_t^4\cdot d\beta/d\log a\cdot (\langle {\rm Pl} \rangle_{N_t}-\langle {\rm Pl} \rangle_{2\cdot N_t})$$
The above method can also be easily generalized for the case of dynamical fermions.

\begin{figure}[h]
\centering
\includegraphics[height=6cm]{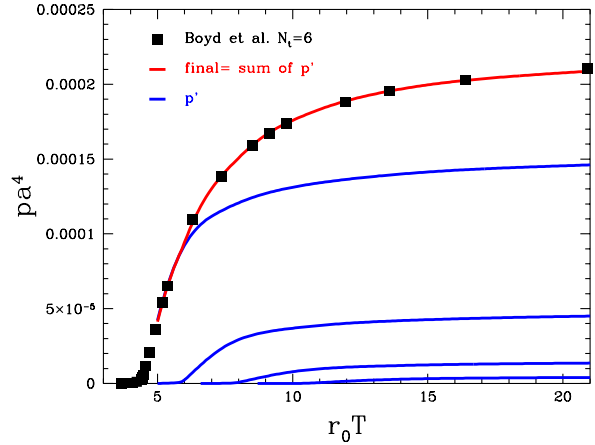}
\caption{The dimensionless pressure as a function of the temperature. By summing up the intermediate terms (indicated by solid blue lines, denoted by $p'$) we arrive at the total renormalized pressure (solid red line), which agrees completely with results obtained by the usual integral method.}
\label{fig:compare}
\end{figure}

We can demonstrate the applicability of this technique by using only the finite temperature $N_t=6$ and $N_t=8$ data of~\cite{Boyd:1996bx} to reproduce their results obtained by the standard method (which uses the more expensive $T=0$ data). Here we will use the ratio $x=6/8$ instead of the previously shown $1/2$. We keep only four terms in the sum (\ref{eq:p_sum}), since the forthcoming terms are practically negligible:
\begin{equation}
\frac{p_{\rm ren}}{T^4}=\bar{p}(T)+x^4 \cdot \bar{p}(xT)+\overbrace{x^8 \cdot \bar{p}(x^2T)}^{p'}+x^{12} \cdot \bar{p}(x^3T)
\end{equation}
On figure \ref{fig:compare}. we show the results obtained by the standard renormalization procedure, and also those by our method.

\section{Setting the scale}

\begin{figure}[h]
\centering
\includegraphics[height=6cm]{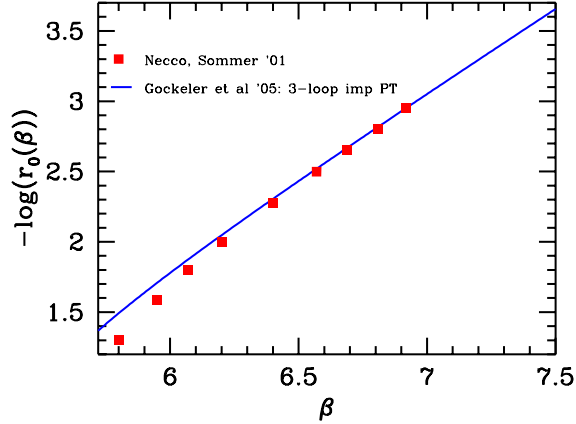}
\caption{Asymptotic scaling seems to be realized, as the improved perturbation theory formula fits the lattice results of the Sommer parameter for larger values of $\beta$.}
\label{fig:scaling}
\vspace*{-0.4cm}
\end{figure}

Our new method is not only capable of reconstructing results of the standard procedure, but it also has a clear advantage over that, since $T=0$ lattices are not needed for the subtraction, only for scale determination. So we need large statistics only for $T>0$ simulations, which are less demanding in terms of computer resources (memory, CPU time). If we wanted to reach very high temperatures, we still needed large, and therefore expensive zero temperature lattices to set the scale. Alternatively, in the asymptotic scaling region, one might determine the scale using improved perturbation theory. To verify that we are indeed in the asymptotic scaling region, we compare results of the Sommer parameter from~\cite{Necco:2001xg}, and a 3-loop improved perturbation theory formula~\cite{Gockeler:2005rv} inserted into the expression of the lattice spacing $a=\Lambda^{-1}(a\to 0) f_{\rm imp PT}(\beta)$. This comparison on figure \ref{fig:scaling} shows that the last few lattice simulation points are in the asymptotic scaling regime, which means that an extrapolation with the perturbative formula might be trusted.

\section{Results I.}

\begin{figure}[h]
\centering
\includegraphics*[height=6cm]{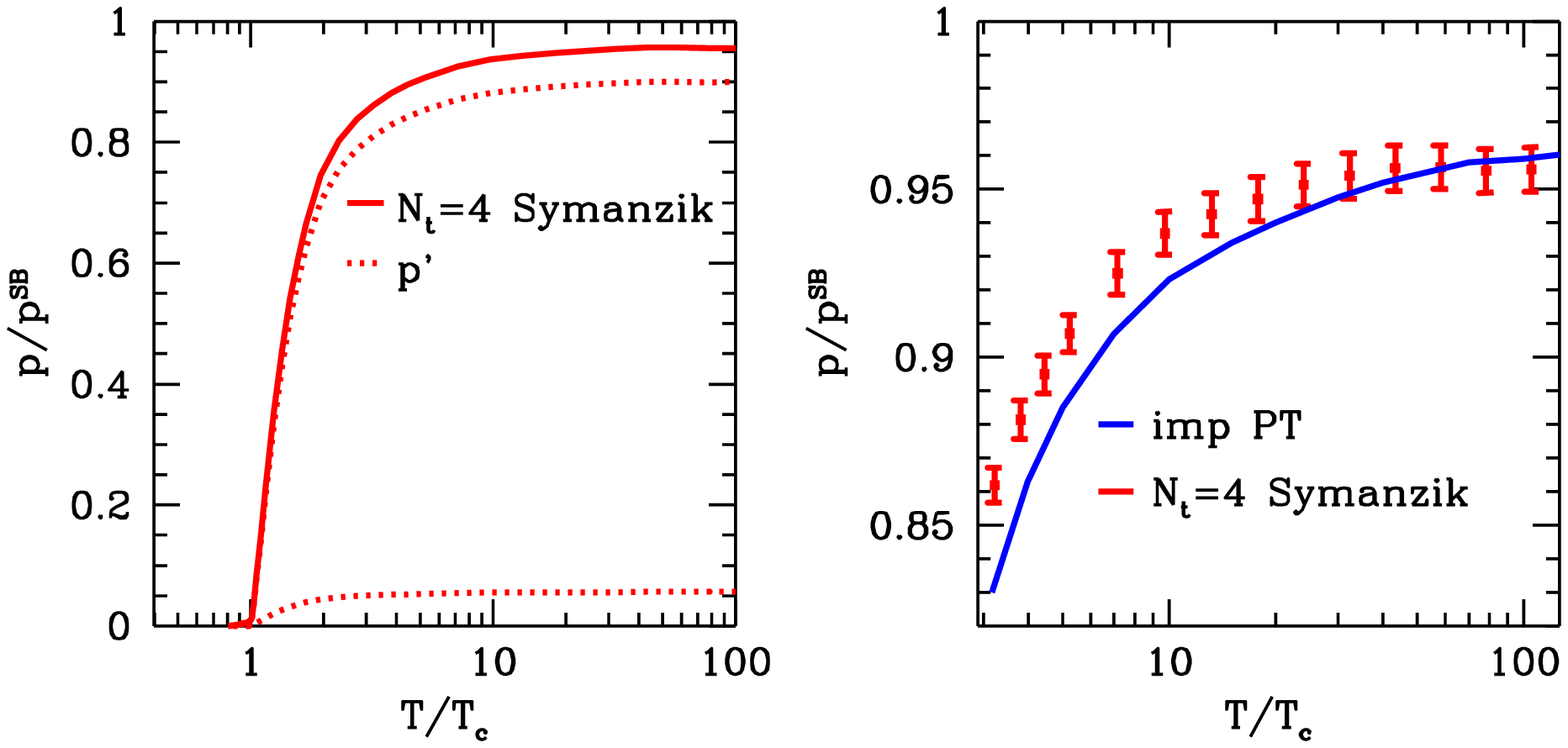}
\caption{We show the two, non-negligible terms (denoted by dotted lines) of the normalized pressure, and their sum (solid line). $P_{SB}$ denotes the pressure of the non-interacting gluon gas.}
\label{fig:p_sym}
\end{figure}

For our simulations we used tree level improved Symanzik action, and an overrelaxation-heatbath algorithm. The simulations were carried out on lattices with temporal extent $N_t=4$; we performed the subtraction on $N_t=8$ lattices. In order to account for decreasing screening masses, we used a rather large aspect ratio of $N_s/N_t = 8$. On figure \ref{fig:p_sym}. we show our results on the pressure.

It is worth mentioning here, that independently of the renormalization procedure, a problem emerges within the integral method framework. Since strictly speaking, the pressure is only exactly zero at $T=0$, in principle we would have to carry out the integration starting from zero temperature. Due to the uncertainty of setting the lower point of the integral, and also due to larger statistical fluctuations in the low temperature regime, we needed about 6 times more statistics for the region below $T_C$, as we did for $T=(1 \ldots 100) \cdot T_C$.

\section{Direct approach}

In the following, we will present a method, which is suitable for measuring the pressure directly, instead of integrating the difference of two, separately measured plaquette variables. This way we can get rid of the above mentioned lower point-related uncertainty in the integral: we can provide a reference point in the pressure.

\begin{figure}[h]
\vspace*{-0.5cm}
\centering
\includegraphics[height=5.8cm]{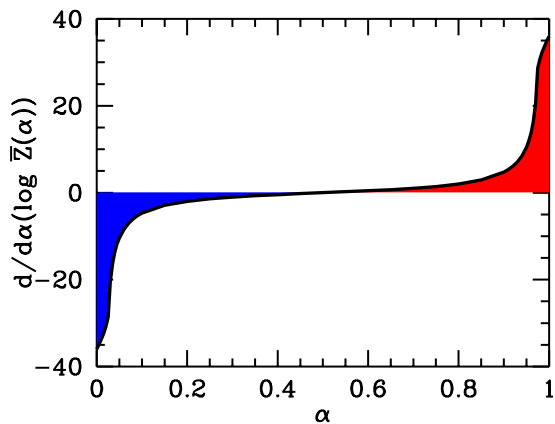}
\caption{Cancellation effect in calculating the pressure by our new method. These results were obtained on $N_t=4$ lattices at $\beta=50$ (which roughly corresponds to the Planck temperature).}
\label{fig:cancel}
\end{figure}

Let us consider one single term of (\ref{eq:p_sum}), namely $\bar p$
as defined by eq. (\ref{eq:p_bar}):

\begin{equation}
\bar{p}=\frac{1}{N_tN_s^3}\log Z(N_t) - \frac{1}{2N_tN_s^3}\log Z(2N_t)=
\frac{1}{2N_tN_s^3}\log \left( \frac{Z(N_t)^2}{Z(2N_t)}\right)
\end{equation}
where we expressed the pressure as $\log Z$. Now schematically we can draw the ratio of the two partition functions like the following (let $S_{2b}$ be the action for the boundary condition in the numerator, and $S_{1b}$ for the one in the denominator):

\begin{figure}[h]
\vspace*{-0.4cm}
\centering
\includegraphics[height=3cm]{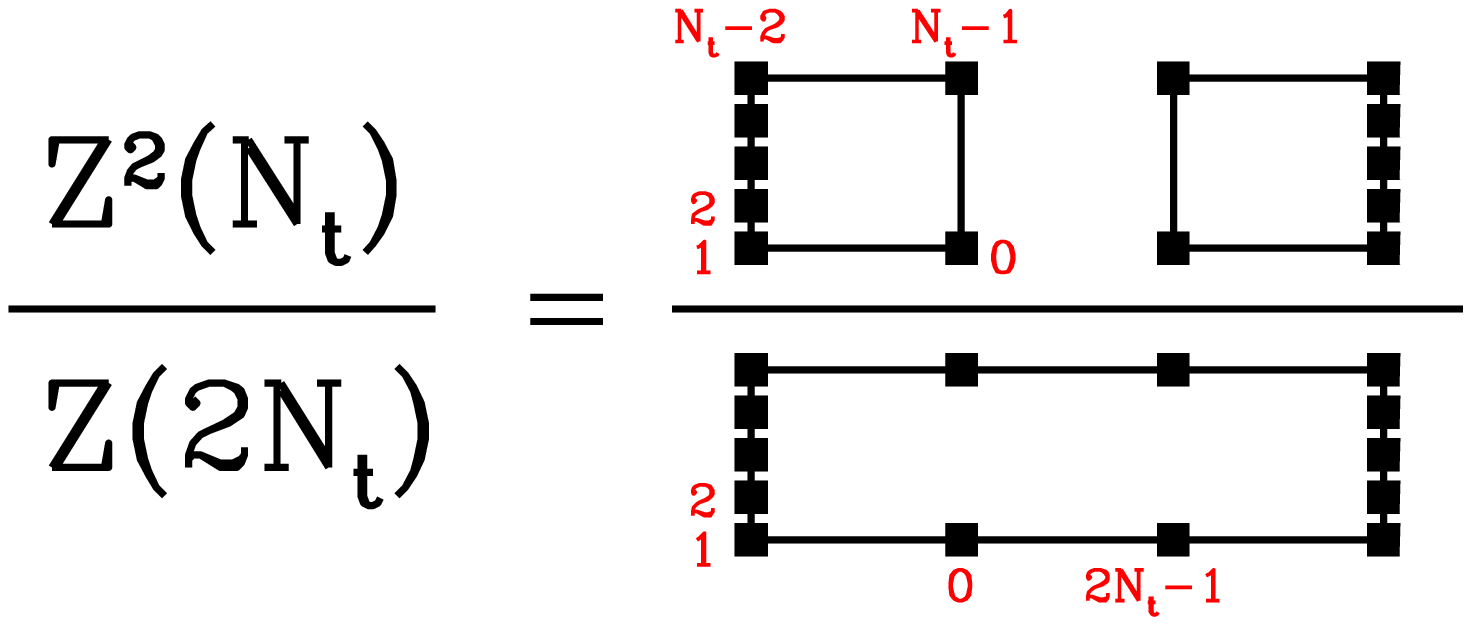}
\vspace*{-0.3cm}
\end{figure}
\noindent
Now let us take an interpolating partition function $\bar Z(\alpha)=\int \mathcal{D}U exp(-(\alpha\cdot S_{2b}+(1-\alpha) \cdot S_{1b}))$, which could be depicted like:

\begin{figure}[h]
\vspace*{-0.3cm}
\centering
\includegraphics[height=1.5cm]{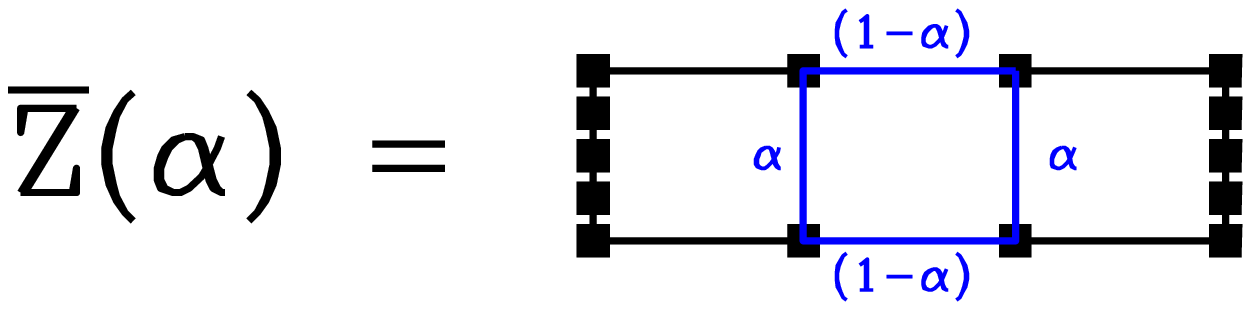}
\vspace*{-0.3cm}
\end{figure}
\noindent
Using $\bar Z (\alpha)$, one obtains

\begin{equation}
\bar{p} \sim \log \left( \frac{Z(N_t)^2}{Z(2N_t)}\right)=\log \left( \frac{\bar{Z}(1)}{\bar{Z}(0)}\right)=
\int_0^1 d\alpha \frac{d \log \bar{Z}(\alpha)}{d\alpha}=
\int_0^1 d\alpha \langle S_{1b} - S_{2b}\rangle 
\label{eq:alpha}
\end{equation}
So we can calculate the pressure itself at any given temperature, without carrying out simulations at lower temperatures, and then performing an integral. However, the cancellation of the plaquette variables, which makes the former method a difficult task, also emerges in this new method. On figure \ref{fig:cancel}., we plot the integrand of (\ref{eq:alpha}) as a function of $\alpha$ to show why it is hard to calculate the integral with a given precision.

Still, it is worth using this new method, since our integrand here is proportional to $N_s^3$, so it is expected to scale with $N_t^{-3}$ instead of $N_t^{-4}$. This means that we gain a factor of $N_t$ with respect to the standard method. In fact, we have to perform the integral in $\alpha$ for each $\beta$ we need. However, there is no need to worry about the lower part of the temperature interval anymore, since using this method we are able to set the integration constant in the pressure.

\section{Results II.}

\begin{figure}[h]
\vspace*{-0.3cm}
\centering
\includegraphics[height=6cm]{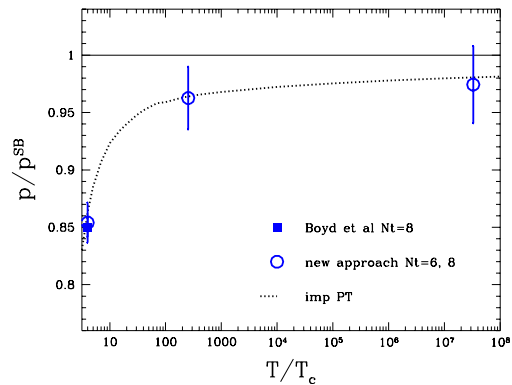}
\vspace*{-0.5cm}
\caption{The pressure, normalized to its Stefan-Boltzmann value, as a function of the temperature obtained by our new techique. Results with smaller discretization errors ($N_t=8$, blue circles), seem to fit improved perturbation theory, and also reproduce results obtained by the standard method at lower temperatures. At the highest temperature, $3 \cdot 10^7 \cdot T_C$, the pressure (within its statistical uncertainty) is consistent with the Stefan-Boltzmann limit.}
\label{fig:p_new}
\end{figure}

The method discussed above gives us the possibility to measure the pressure at very high temperatures. This was carried out using lattices with temporal extension $N_t=4$, $N_t=6$ and $N_t=8$. The temperature interval ranged from $4 \cdot T_C$ upto $3 \cdot 10^7 \cdot T_C$. Our results are shown on figure \ref{fig:p_new}. A comparison can be done with the standard method, namely the result of~\cite{Boyd:1996bx} for smaller temperatures. Here we present the results which are closest to the continuum limit, thus $N_t=8$. These results nicely follow the perturtbative predictions (see~\cite{Braaten:1995jr, Kajantie:2002wa, Laine:2006cp}, and also~\cite{Blaizot:2003tw, Andersen:2002ey}). However, more statistics is needed to determine the whole applicability region of the perturbative approach. It is important to point out, that at the highest temperature point, the pressure (within its statistical uncertainty) is already consistent with the Stefan-Boltzmann limit.

\section{Conclusion}

We presented two new methods to determine the equation of state for QCD. Compared to previous techniques, these methods allow to extend the temperature range by orders of magnitude. We developed a renormalization procedure of the pressure, for which no $T=0$ simulations are needed, and thus is less expensive than the methods, which were used previously. We also presented a new approach, which enables us to measure the pressure directly. Using this approach we gain a factor of $N_t$ in CPU time with respect to the standard method. Our approach can also be used to set the integration constant, which provides a reference point for the pressure. Based on these new methods we presented first results on the pressure of the pure gauge theory at temperatures that could not have been reached before.

\paragraph{Acknowledgements}

This research was supported by the EU under grant I3HP and DFG under grant FO 502/1. We would like to acknowledge helpful discussions with M.~Laine and Y.~Schr\"{o}der.

\end{document}